\documentclass[final,authoryear,3p,times]{elsarticle}

\usepackage{graphicx,amssymb,amsmath,amsthm,bm,accents,color,subfig,natbib,hyperref}

\journal{Int. J. Eng. Sci.}

\begin{document}

\begin{frontmatter}

\title{Comments on: ``Starting solutions for some unsteady unidirectional flows of a second grade fluid,'' [Int.\ J.\ Eng.\ Sci.\ 43 (2005) 781]}

\author[NU]{Ivan C. Christov\fnref{ivan}}
\ead{christov@alum.mit.edu}
\ead[url]{http://alum.mit.edu/www/christov}

\author[ERC]{P. M. Jordan}

\fntext[ivan]{Corresponding author.}
\address[NU]{Department of Engineering Sciences and Applied Mathematics, Northwestern University, Evanston, IL 60208, USA}

\address[ERC]{Entropy Reversal Consultants (L.L.C.), P.\ O.\ Box 691, Abita Springs, LA 70420, USA}

\begin{abstract}
A significant mathematical error is identified and corrected in a recent highly-cited paper on oscillatory flows of second-grade fluids [Fetecau \& Fetecau (2005). {\it Int.\ J.\ Eng.\ Sci., 43}, 781--789]. The corrected solutions are shown to agree identically with numerical ones generated by a finite-difference scheme, while the original ones of Fetecau \& Fetecau do not. A list of other recent papers in the literature that commit the error corrected in this Comment is compiled. Finally, a summary of related erroneous papers in this journal is presented as an Appendix.
\end{abstract}


\end{frontmatter}

\section{Introduction}
\label{sec:intro}
\citet{FF05} considered the transient version of Stokes' second problem for a second-grade (SG) fluid. Unfortunately, due to a mathematical error in imposing the boundary condition, an incorrect and unphysical solution is obtained. The purpose of the present Comment is to correct this error, which has unfortunately been promulgated through scores of papers, and to offer a new perspective on the subject.

In the literature, many errors (of increasing variety and frequency) have been made by a growing list of authors  attempting to obtain solutions to initial-boundary-value problems arising from  simple flows of non-Newtonian fluids. For in-depth discussions (including identification of the mistakes and how to correct them) in a variety of contexts see, e.g., \citep{JPB04,JP04,J05,P08,P08b,CJ09,R09,J10,CC10,C10,AK10,L10,S10,MMA11,C11,K12}.

Another unsuccessful attempt, concurrent with that of \citet{FF05}, at solving the transient version of Stokes' second problem for a SG fluid can be found in \citep{ANHH06}. We shall not concern ourselves with the latter work in the discussion below because, it suffices to note, \citet{ANHH06} solve a {linear constant-coefficient} second-order ordinary differential equation by a perturbation series for which they take a \emph{dimensional} quantity as the small parameter without giving a point of reference.

\section{Start-up problem for a half-space}
\label{sec:half-space}
\citet{FF05} first attempt to solve the following initial-boundary-value problem (IBVP):
\begin{subequations}\begin{align}
\frac{\partial u}{\partial t} &= \nu \frac{\partial^2 u}{\partial y^2} + \alpha \frac{\partial^3 u}{\partial y\partial t\partial y}, &&(y,t)\in(0,\infty)\times(0,\infty),\label{eq:pde}\\
u(0,t) &= U_0f(\omega t)H(t), &&t >0, \label{eq:bc_0}\\
u(y,t)&\to 0\quad\text{as}\quad y\to\infty, &&t>0, \label{eq:bc_1}\\
u(y,0) &= 0, &&y>0.
\end{align}\label{eq:ibvp}\end{subequations}
Here, we break with the notation of \citet{FF05} because of the similarity of the symbols for viscosity and velocity employed therein. To this end, $\nu$ is the kinematic viscosity, $\alpha$ is the ratio of the first material modulus of the SG fluid to its density, $u=u(y,t)$ is the fluid velocity in the $x$-direction, $U_0$ and $\omega$ are, respectively, the amplitude and the frequency of the oscillations of the infinite plate upon which this SG fluid rests. Two cases are considered: $f(\cdot)=\cos(\cdot)$ and $f(\cdot) = \sin(\cdot)$.

The Heaviside unit step function $H(t)$ multiplying $U_0f(\omega t)$ in Eq.~\eqref{eq:bc_0} is neglected by \cite{FF05}. \citet[p.~101--102]{S51}, though lacking the mathematical tools such as the Heaviside function or the operational calculus, was able to clearly (and elloquently) explain the difference between imposing  $u(0,t) = U_0f(\omega t)H(t)$ and imposing $u(0,t) = U_0f(\omega t)$ for a Newtonian fluid. Hence, though it is claimed that the problem in which ``At time $t = 0^+$ the rigid plate begins to oscillate.'' \citep[p.~783]{FF05} is solved, this is not the case. In fact, it is unclear whether the expressions derived by \citet{FF05} solve \emph{any} physically relevant problem; for Stokes' first problem for a SG fluid, the similarly incorrect solution could be re-interpreted as one arising from a fictitious body force or a different, but physical, boundary condition \citep{CC10}.

Upon applying the Fourier sine transform to Eq.~\eqref{eq:pde}, the second term on the right-hand side becomes
\begin{equation}
\frac{\partial}{\partial t}\int_0^\infty \frac{\partial^2 u}{\partial y^2}(y,t) \sin(\xi y) \,\mathrm{d}y = \frac{\partial}{\partial t} \left[-\xi^2 u_{\mathrm{s}}(\xi,t) + \xi u(0,t)\right] = -\xi^2\frac{\partial u_{\mathrm{s}}}{\partial t}(\xi,t) + \xi \frac{\partial u}{\partial t}(0,t),
\end{equation}
where the first equality follows from assuming that $u$ is twice continuously differentiable with respect to $y$ on $(0,\infty)$, and that $u$, $\partial u/\partial y$ and $\partial^2 u/\partial y^2$ are integrable functions of $y$ on $(0,\infty)$. Note that the decay condition stated in Eq.~\eqref{eq:bc_1} is short-hand notation for this. Here, an ``s'' subscript denotes the image of a function in the Fourier sine transform domain, and $\xi$ denotes the Fourier sine transform parameter. We have employed the Fourier sine transform convention of \citet{CB78}.

\subsection{Case I: $f(\cdot) = \cos(\cdot)$.}\label{sec:cos}

Using the Schwartz--Sobolev theory of \emph{distributions}, also known as \emph{generalized functions} \citep[see, e.g.,][Chap.~2]{K83}, we have, from Eq.~\eqref{eq:bc_0}, that 
\begin{equation}
\frac{\partial u}{\partial t}(0,t) = U_0f(\omega t)\delta(t) + \omega U_0f'(\omega t)H(t) = U_0\cos(\omega t)\delta(t) - \omega U_0\sin(\omega t)H(t)
\label{eq:plate-accl}
\end{equation}
for this boundary condition, where $\delta(\cdot)$ is the Dirac delta distribution. By comparing Eq.~\eqref{eq:plate-accl} to the right-hand side of \citep[Eq.~(3.4)]{FF05}, it is apparent that the error committed by the latter authors boils down to taking $\delta(t) = 0$ and $H(t)=1$.  However, this is both physically and mathematically wrong as $\delta(\cdot)$ and $H(\cdot)$ are distributions, hence \emph{they do not have point values}, and they are certainly \emph{not} equal to 0 and 1, respectively, for all $t$. [A similar incorrect claim may be found in \citep[Eqs.~(3) and (4)]{TM09}.] The equality in Eq.~\eqref{eq:plate-accl} is {not} understood in a pointwise sense for individual values of $t$, rather it is understood as being true upon multiplication of both sides by a smooth function and integration over $t\in(0,\infty)$.

Another way to see why \citep[Eqs.~(3) and (4)]{TM09}, which can be considered as the reasoning behind the mistake made by \citet{FF05}, are incorrect is to perform the following thought experiment. Let us specify, and we are free to do so, that the plate is set in motion at some $t=t_0^+$, where $t_0 > 0$, while being at rest for all $t \le t_0$. Then, all terms involving the Heaviside and Dirac distributions that appear in this new IBVP (whose solution we shall denote by $\tilde{u}$) are of the form $H(t-t_0)$ and $\delta(t-t_0)$ with $t\in(0,\infty)$ as before. Since we require $t_0>0$, it should now be even more obvious that the claim $\delta(t-t_0) = 0$ and $H(t-t_0)=1$ for all $t\in(0,\infty)$ is incorrect. Finally, from the change of variable $t\mapsto t+t_0$, it follows that $u(y,t) = \tilde{u}(y,t+t_0)$ satisfies the original IBVP~\eqref{eq:ibvp}, demonstrating that the claims made by \citet{TM09} (and equivalent manipulations by \citet{FF05}) are erroneous.

Completing the application of the Fourier sine transform in $y$ and then applying the Laplace transform in $t$, the solution to the subsidiary equation in the Fourier--Laplace domain is
\begin{equation}
\overline{u_{\mathrm{s}}}(\xi,s) = \frac{U_0\xi}{s(1+\alpha\xi^2) + \nu\xi^2}\left(\frac{\nu s - \alpha\omega^2}{s^2+\omega^2} + \alpha \right).
\label{eq:subsidiary}
\end{equation}
Note that this expression (correctly) takes into account all information given in Eq.~\eqref{eq:ibvp}. Here, a bar over a quantity denotes its image in the Laplace transform domain and $s$ denotes the Laplace transform parameter.

Inverting Eq.~\eqref{eq:subsidiary} using partial fractions, a standard table of Laplace inverses and the definition of the Fourier sine transform, we arrive at
\begin{multline}
u(y,t) = U_0H(t)\left(\frac{2}{\pi}\int_0^\infty \frac{\xi\sin(y\xi)}{\nu^2\xi^4 + \omega^2(1 + 
 \alpha\xi^2)^2}\Bigg\{ -\frac{\nu^2\xi^2}{1 + \alpha\xi^2}\exp\left(-\frac{\nu\xi^2}{1 + \alpha\xi^2}t\right)\right. \\ + \left. \nu\omega \sin(\omega t) + \left[\nu^2\xi^2 + \alpha \omega^2 (1 + \alpha\xi^2)\right] \cos(\omega t) \Bigg\}  \,\mathrm{d}\xi \right).
\label{eq:soln-cos}
\end{multline}
We leave it to the reader to find the correct \emph{long-time} (post-transient) expression for the solution. Here it is important to note that this limiting form of the solution is \emph{not} independent of time (i.e., it is not, strictly speaking, a ``steady state,'' as claimed in \citep{FF05}).

\subsection{Case II: $f(\cdot) = \sin(\cdot)$.}
Now, Eq.~\eqref{eq:plate-accl} becomes
\begin{equation}
\frac{\partial u}{\partial t}(0,t) = U_0\sin(\omega t)\delta(t) + \omega U_0\cos(\omega t)H(t).
\end{equation}
In this case, however, there is a ``self-canceling error,'' as discussed in \citep{CC10,C10}, because $\mathcal{L}\{\sin(\omega t)\delta(t)\} = 0$. In other words, the term in the subsidiary equation that \citet{FF05} neglected does not contribute to the solution. 

For this boundary condition, the solution of the subsidiary equation in the Fourier--Laplace domain is
\begin{equation}
\overline{u_{\mathrm{s}}}(\xi,s) = \frac{U_0\xi}{s(1+\alpha\xi^2) + \nu\xi^2}\left(\frac{\nu\omega + \alpha \omega s}{s^2+\omega^2} \right).
\label{eq:subsidiary2}
\end{equation}
Inverting Eq.~\eqref{eq:subsidiary2} using partial fractions, a standard table of Laplace inverses and the definition of the Fourier sine transform, we arrive at
\begin{multline}
u(y,t) = U_0H(t)\Bigg( \frac{2}{\pi} \int_0^\infty \frac{\xi\sin(y\xi)}{\nu^2\xi^4 + \omega^2(1+ \alpha\xi^2)^2}\Bigg\{ \nu\omega\exp\left(-\frac{\nu\xi^2}{1 + \alpha\xi^2}t\right) \\ 
 - \nu\omega \cos(\omega t) + \left[\nu^2 \xi^2 + \alpha \omega^2 (1 + \alpha\xi^2) \right] \sin(\omega t) \Bigg\}\,\mathrm{d}\xi \Bigg).
\label{eq:soln-sin}
\end{multline}
Despite the ``self-canceling error,'' we give this solution for completeness because it differs from \citep[Eq.~(3.9)]{FF05}, perhaps due to a choice made during the inversion procedure.

\section{Start-up problem for a strip}
\label{sec:strip}
Next, we turn to the problem considered in \citep[Sect.~4]{FF05}. Restricting Eq.~\eqref{eq:ibvp} to the strip $0\le y \le d$, we obtain a new IBVP:
\begin{subequations}\begin{align}
\frac{\partial u}{\partial t} &= \nu \frac{\partial^2 u}{\partial y^2} + \alpha \frac{\partial^3 u}{\partial y\partial t\partial y}, &&(y,t)\in(0,\infty)\times(0,d),\label{eq:pde_strip}\\
u(0,t) &= U_0f(\omega t)H(t), &&t >0, \label{eq:bc_0_strip}\\
u(d,t) &= 0, &&t > 0,\label{eq:bc_1_strip}\\
u(y,0) &= 0, &&y\in(0,d).\label{eq:ic_strip}
\end{align}\label{eq:ibvp_strip}\end{subequations}
Rather than follow the approach in \citep[Sect.~4]{FF05}, here we make use of a significantly more elegant solution procedure involving only the Laplace transform in time only as suggested by  \citet{J05}.

An eigenfunction expansion for a start-up problem (as attempted by \citet{FF05} and some of the papers discussed in the Appendix of this Comment) presents many opportunities for making a mistake. If said eigenfunction expansion were performed correctly, it would produce an identical solution to the one obtained by the more robust method we now employ. To this end, applying only the Laplace transform in $t$ to Eqs.~\eqref{eq:pde_strip}--\eqref{eq:bc_1_strip}, using the initial condition in Eq.~\eqref{eq:ic_strip}, and then solving the resulting subsidiary equation, we obtain
\begin{equation}\label{Couette_TDS}
\overline{u}(y,s)=\frac{U_0\sinh\left[\left(1-ly/\sqrt{\alpha}\right)\sqrt{s/(l^2\kappa + sl^2)}\right]}{\sinh\left[\sqrt{s/(l^2\kappa + sl^2)}\right]}
\times\left\{\begin{aligned}
\displaystyle{\frac{s}{s^2+\omega^2}}, &\quad f(\cdot) =\cos(\cdot)\\
\displaystyle{\frac{\omega}{s^2+\omega^2}}, &\quad f(\cdot)=\sin(\cdot)
\end{aligned}\right\},
\end{equation}
where we have introduced the dimensionless parameter $l := d^{-1}\sqrt{\alpha}$ and defined $\kappa :=\nu/\alpha$ for convenience.
 
To make the expression for the solution more tractable, we introduce the dimensionless independent variables $\tilde y = y/\sqrt{\alpha}$ and $\tilde t = \kappa t$ together with the dimensionless frequency $\Omega := \omega/\kappa$. Now, it is readily established from \citep[Eq.~(3.4)]{J05} and the  convolution theorem for the Laplace transform, that the exact solution of IBVP~\eqref{eq:ibvp_strip} for the case $f(\cdot) =\sin(\cdot)$ is given by
\begin{equation}\label{Couette_sin_sol}
u(\tilde y,\tilde t) = U_{0}H(\tilde t)\left\{\sin(\omega t)\left(1 - l\tilde y\right)
+\frac{2\Omega}{\pi}\sum_{n=1}^{\infty}
\left[\frac{\sigma_n\exp(-\sigma_n\tilde t) - \sigma_n\cos(\Omega \tilde t) - \Omega\sin(\Omega \tilde t)}{\sigma_n^2 + \Omega^2}\right]\frac{\sin(l n\pi \tilde y)}{n(1+l^{2}n^2\pi^2)} \right\},
\end{equation}
where $\sigma_n := l^2n^2\pi^2(1+l^{2}n^2\pi^2)^{-1}$. In turn, using Eq.~\eqref{Couette_sin_sol}, the fact that $\mathcal{L}^{-1}\{s\}=\delta'(t)$, and, once again, the convolution theorem for the Laplace transform, the exact solution of IBVP~\eqref{eq:ibvp_strip} for the case $f(\cdot) =\cos(\cdot)$ is found to be
\begin{equation}\label{Couette_cos_sol}
u(\tilde y,\tilde t) = U_{0}H(\tilde t)\left\{\cos(\omega t)\left(1 - l\tilde y \right)
-\frac{2}{\pi}\sum_{n=1}^{\infty}
\left[\frac{\sigma_n^2\exp(-\sigma_n\tilde t) - \sigma_n\Omega\sin(\Omega \tilde t) + \Omega^2 \cos(\Omega \tilde t)}{\sigma_n^2 + \Omega^2}\right]\frac{\sin(l n\pi \tilde y)}{n(1+l^{2}n^2\pi^2)} \right\}.
\end{equation}

Finally, we refer the reader to \citet{J05} for a discussion of the equivalent error for the case of Stokes' first problem (see also Items \ref{it:1} and \ref{it:2} in the Appendix).

\section{Illustrating examples}
\label{sec:illustrated}
Using the unconditionally stable and formally second-order accurate finite-difference scheme of \citet{CC10}, we solve the IBVPs \eqref{eq:ibvp} and \eqref{eq:ibvp_strip} numerically and compare the resulting solutions to the wrong transform solutions of \citet{FF05} and the corrected transform solutions presented in Sects.~\ref{sec:half-space} and \ref{sec:strip} above. The scheme parameters are as given in \citep[Sect.~3]{CC10}. The integral representations of the analytical solutions are evaluated using the high-precision numerical integration routine {\tt NIntegrate} of the software package {\sc Mathematica} (ver.~7.0.1). 

\begin{figure}[!ht]
\centering
\subfloat[$\tilde t = 1$]{\includegraphics[width=0.45\textwidth]{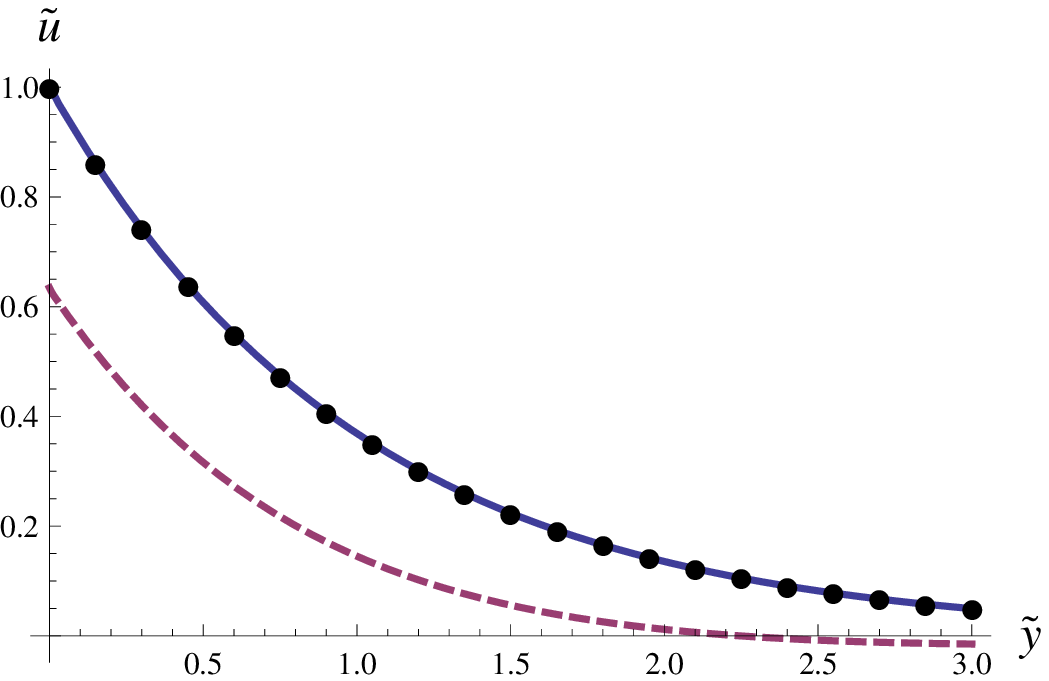}}\hspace{10mm}
\subfloat[$\tilde t = 3.5$]{\includegraphics[width=0.45\textwidth]{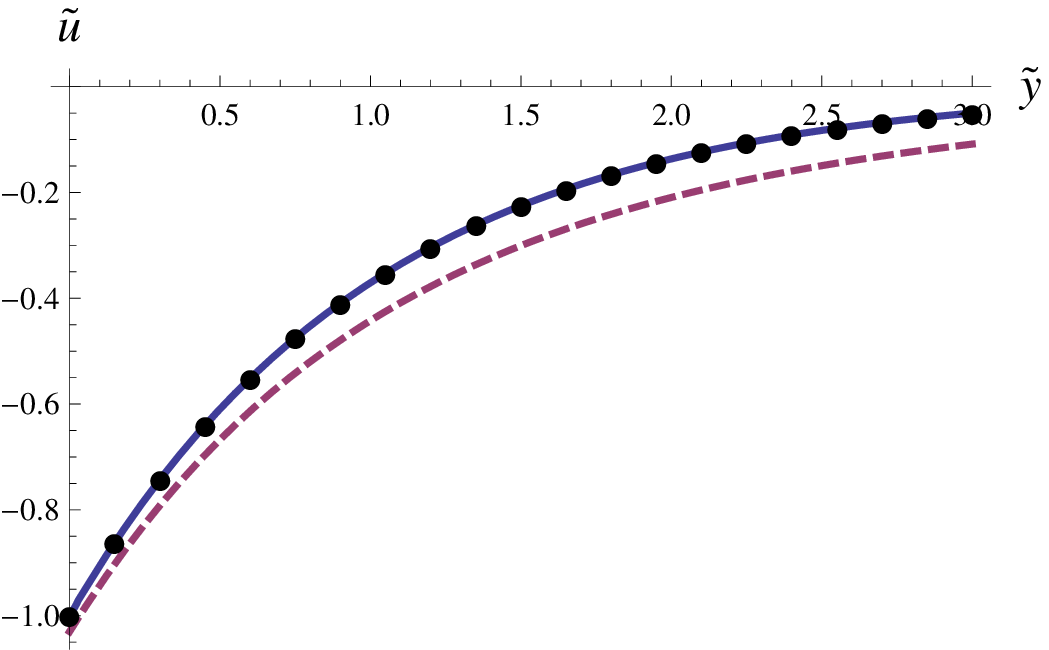}}
\caption{(Color online.) $\tilde u$ vs.\ $\tilde y$ for the half-plane problem with $f(\cdot) = \cos(\cdot)$ and $\Omega = 2\pi$. Dashed: incorrect solution \citep[Eq.~(3.6)]{FF05}; solid: correct solution [Eq.~\eqref{eq:soln-cos} above]; dots: numerical solution [using the scheme in \citep[Eq.~(4)]{CC10}].}
\label{fig:soln-cos}
\end{figure}

\begin{figure}[!ht]
\centering
\subfloat[$\tilde t = 1$]{\includegraphics[width=0.45\textwidth]{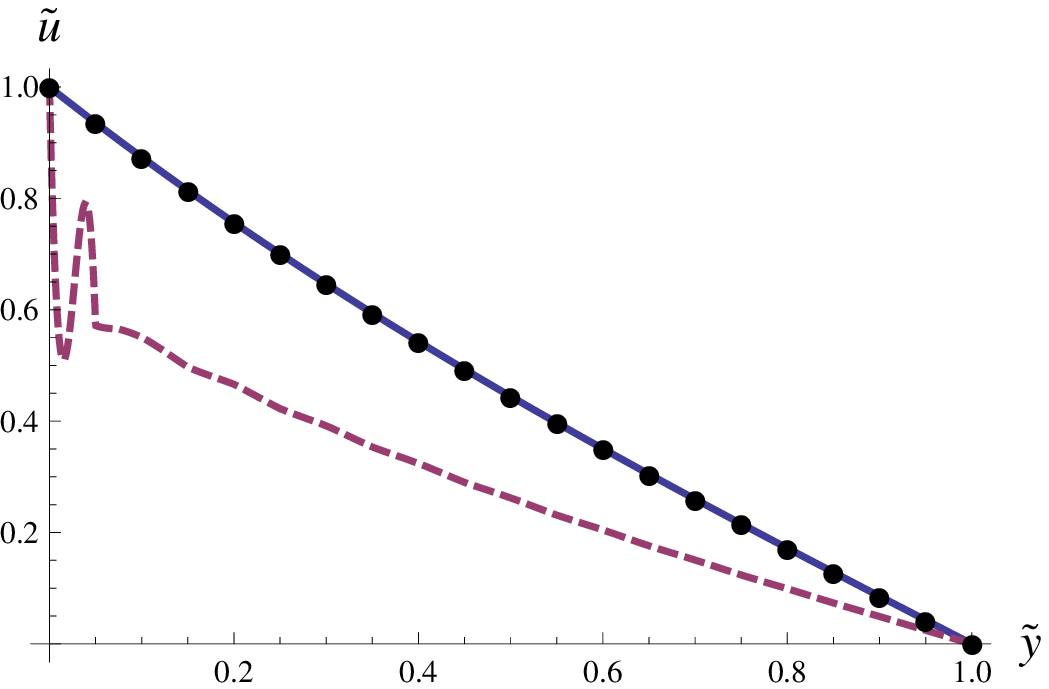}}\hspace{10mm}
\subfloat[$\tilde t = 2.5$]{\includegraphics[width=0.45\textwidth]{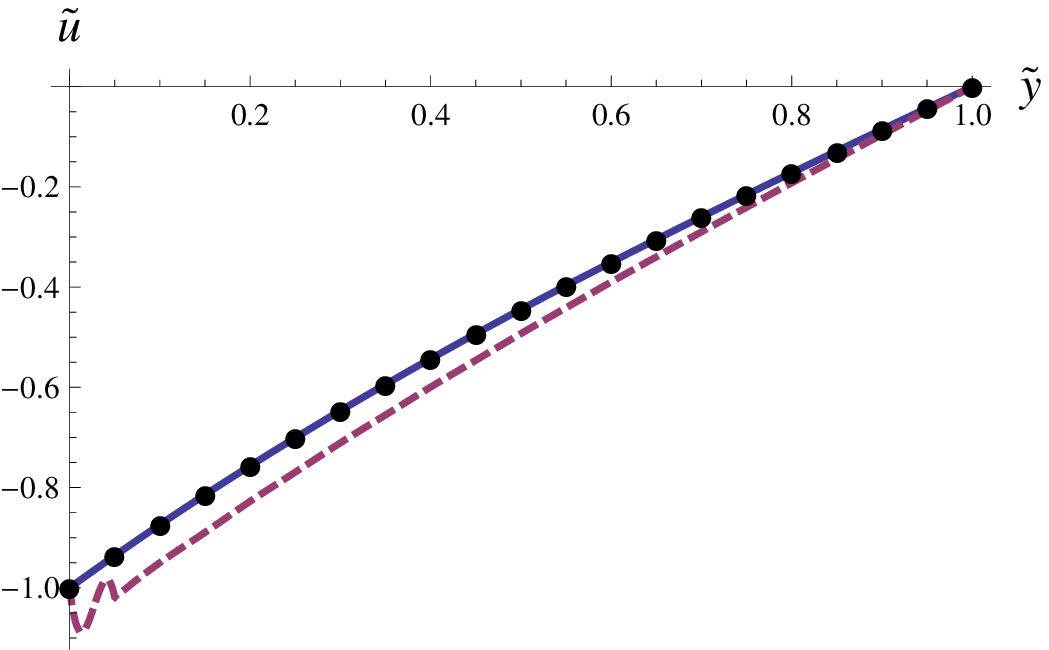}}
\caption{(Color online.) $\tilde u$ vs.\ $\tilde y$ for the strip problem with $l=1$, $f(\cdot) = \cos(\cdot)$ and $\Omega = 2\pi$. Dashed: 100 terms of the incorrect series solution \citep[Eq.~(4.6)]{FF05}; solid: 100 terms of the correct solution [Eq.~\eqref{Couette_cos_sol} above]; dots: numerical solution [using the scheme in \citep[Eq.~(4)]{CC10}].}
\label{fig:strip-cos}
\end{figure}

Figure~\ref{fig:soln-cos} shows a comparison of the erroneous, corrected and numerical solutions  for the half-space problem with $f(\cdot)=\cos(\cdot)$. Similarly, Fig.~\ref{fig:strip-cos} shows a comparison for the problem on a strip with $f(\cdot)=\cos(\cdot)$. We do not give plots for the case of $f(\cdot) = \sin(\cdot)$ because the aforementioned self-canceling error renders the solution curves visually indistinguishable.

Here, it is important to note another problem with the work of \citet{FF05}: when presenting numerical values of physical quantities units are neglected; additionally, values of questionable relevance are chosen for the material parameters. For example, for the purposes of \citep[Fig.~1]{FF05}, the viscosity and density of the fluid are taken to be those of glycerin, which (under normal conditions) cannot be claimed to be a SG fluid without justification. Arguably, it is impossible to pick a realistic value for $\alpha$ because undisputed experimental measurements do not exist \citep{DR95}. However, for the purposes of making a plot of the solution, we can scale both $\nu$ and $\alpha$ out by using the \emph{dimensionless} variables $\tilde y = y / \sqrt{\alpha}$, $\tilde t = t(\nu/\alpha)$, $\Omega = \omega (\alpha/\nu)$ and $\tilde u = u/U_0$ \citep{BRG95}. 

From Fig.~\ref{fig:soln-cos}, it is clear that the solution from Eq.~\eqref{eq:soln-cos} agrees identically with the corresponding numerical solution. Meanwhile, the incorrect solution \citep[Eqs.~(3.6)]{FF05} does not even satisfy the boundary condition $\lim_{y\to0^+} u(y,t) = U_0$ $(t > 0)$. Unlike the case of Stokes' first problem for a SG fluid corrected by \citet{CC10}, here it is unclear what kind of boundary condition the wrong solution satisfies, or whether it has any physical meaning. Also, note the erroneous prediction, by the solution of \citet{FF05}, of a \emph{back-flow} for $\tilde y \gtrsim 2.25$, preventing the solution from satisfying the asymptotic boundary condition in Eq.~\eqref{eq:bc_1}.\footnote{\cite{P08,P08b} uncovered a number of papers on steady boundary layer flows of non-Newtonian fluids in which the supposed solution also exhibits this kind of obviously unphysical behavior.} What is more, for the case of a strip, the curve drawn based on \citep[Eq.~(4.6)]{FF05} appears to satisfy the boundary condition but exhibits highly unphysical oscillations near the moving plate. This is similar to the behavior of the incorrect solution to Stokes' first problem on a strip corrected by \citet{J05}.

For all cases considered in this section, the erroneous solutions from \citep{FF05} appear to eventually agree with the correct and numerical solutions in the limit of $\tilde t \gg 1$, though we are unable to confirm the long-time expressions given in \citep{FF05}. In addition, for brevity, we did not check the results in \citep[Sect.~5]{FF05} for correctness. Moreover, any equation in \citep{FF05} that is not explicitly corrected in the discussion above should \emph{not} be assumed to be correct.

\section{Conclusion}

The contribution of the present work is in correcting a number of mistakes made in an attempt to solve the transient version of Stokes' second problem for a SG fluid. Where possible, we have also suggested alternative analytical approaches, which prevent the errors committed by \cite{FF05}. Unfortunately, the very same mistake has already been committed in the following papers: \citet{AFS06,CC06,KSFH07,FHFA08,KNFH08,KAQ09,HHAF10,KAQF10,KAFQ10,AZR10,YL10}.

The present Comment is also useful when faced with a supposed solution to unidirectional flow of a SG, Oldroyd-B or the so-called ``generalized'' SG and Oldroyd-B fluids executing the same motion. Specifically, there is a parameter regime in which the SG fluid, studied herein under the transient version of Stokes' second problem, and studied in \citep{CC10}, under Stokes' first problem, is reproduced. Hence, authors of such studies must be able to show their Fourier--Laplace domain solution agrees (in the proper parameter limit) with, e.g., Eq.~\eqref{eq:subsidiary}, Eq.~\eqref{eq:subsidiary2} or \citep[Eq.~(2)]{CC10}. It is worth noting, however, that not all derivations leading to the latter Fourier--Laplace domain solutions are mathematically correct \citep[see][]{C11}.

\section*{Acknowledgments}
The authors would like to thank Prof.\ C.\ I.\ Christov for his advice and encouragement, and Prof.\ K.\ R.\ Rajagopal for informative discussions. The time-shift argument presented in Sect.~\ref{sec:cos} was suggested by one of the anonymous reviewers.

\bibliographystyle{model5-names}
\bibliography{stokes_2nd_pb,../MRC/stokes_1st_pb}

\appendix
\section*{Appendix. Other erroneous recent papers on simple flows of non-Newtonian fluids in this journal}

\begin{enumerate}

\item\label{it:1} \citet{HAS00} present, in their Sect.~2.3, an incorrect solution for the plane Couette flow of a SG fluid between parallel plates. The correct solution is given by \citet{J05}. \citet{HAS00} have neglected to include $H(t)$ in their initial condition in their  Eq.~(12), which renders the  transformation in their Eq.~(13) erroneous. If the $H(t)$ pre-factor were carried through, there would be an additional term proportional to $\delta(t)$ on the left-hand side of their Eq.~(15). What is more, the initial condition, which must be understood as the limit $t\to0^-$ of the solution \citep{J10}, would require that the right-hand side of their Eq.~(18) is zero, identically, \emph{even after} the attempted transformation. The error in \citep[Sect.~2.3]{HAS00} also impacts a number of results given in their Section 3.

\item\label{it:2} \citet{HAS01} present, in their Sect.~6, incorrect solutions for the plane Couette flow of an Oldroyd-B fluid between parallel plates. The mistake is the same as described in Item \ref{it:1}; again, see \citet{J05,J10}.

\item\label{it:3} \citet{FF06} omit $H(t)$ in their Eq.~(7). As in Item~\ref{it:1} above, this leads to the lack of a term proportional to $\delta(t)$ on the left-hand side of their Eq.~(10.1), and the initial condition in their Eq.~(10.3) is incorrect. Therefore, when the transformation given in their Eq.~(9) is made, it is implicitly assumed the outer cylinder \emph{has been and will continue} to oscillate. Hence, the problem solved is not that of start-up as claimed. 

\item \citet{EI07b} present, in their Sect.~4, incorrect solutions to and incorrect discussion of Stokes' first problem for a SG fluid. These are corrected in \citep{CC10}.

\item \citet{TL05} claim that Sect.~3 of their paper presents ``Poiseuille flow due to a constant pressure gradient.'' However, the start-up flow for a cylinder that begins to rotate at constant angular velocity is considered instead. Unfortunately, $H(t)$ is missing from the boundary condition in their Eq.~(6), which makes the transformation in their Eq.~(7) and the resulting governing Eq.~(8) incorrect. This is the same mistake as in Item \ref{it:3} above.

\item \citet{ZF07} perform an energy analysis using an incorrect solution, rendering their analysis erroneous. The correct solution to be used is given in \citep[Eq.~(3)]{CC10}.

\end{enumerate}

\end{document}